\documentclass[aps,prl,onecolumn,showpacs,preprintnumbers]{revtex4}
\usepackage{amsmath}
\usepackage{dcolumn}
\usepackage{bm}
\usepackage{subfigure}
\usepackage{amsfonts}
\usepackage{amssymb}
\usepackage{makeidx}
\usepackage{epsfig}
\usepackage{graphicx}

\begin{document}

\title{Quantum-wave equation and Heisenberg inequalities of covariant
quantum gravity}
\author{Claudio Cremaschini}
\affiliation{Institute of Physics and Research Center for Theoretical Physics and
Astrophysics, Faculty of Philosophy and Science, Silesian University in
Opava, Bezru\v{c}ovo n\'{a}m.13, CZ-74601 Opava, Czech Republic}
\author{Massimo Tessarotto}
\affiliation{Department of Mathematics and Geosciences, University of Trieste, Via
Valerio 12, 34127 Trieste, Italy}
\affiliation{Institute of Physics, Faculty of Philosophy and Science, Silesian University
in Opava, Bezru\v{c}ovo n\'{a}m.13, CZ-74601 Opava, Czech Republic}
\date{\today }

\begin{abstract}
Key aspects of the manifestly-covariant theory of quantum gravity
(Cremaschini and Tessarotto 2015-2017) are investigated. These refer, first,
to the establishment of the 4-scalar, manifestly-covariant evolution quantum
wave equation, denoted as covariant quantum gravity (CQG) wave equation,
which advances the quantum state $\psi $ associated with a prescribed
background space-time. In this paper, the CQG-wave equation is proved to
follow at once by means of a Hamilton-Jacobi quantization of the classical
variational tensor field $g\equiv \left\{ g_{\mu \nu }\right\} $ and its
conjugate momentum, referred to as (canonical) $g-$\emph{quantization}. The
same equation is also shown to be variational and to follow from a
synchronous variational principle identified here with the \emph{quantum
Hamilton variational principle}. The corresponding quantum hydrodynamic
equations are then obtained upon introducing the Madelung representation for
$\psi $, which provide an equivalent statistical interpretation of the
CQG-wave equation. Finally, the quantum state $\psi $ is proved to fulfill
generalized Heisenberg inequalities, relating the statistical measurement
errors of quantum observables. These are shown to be represented in terms of
the standard deviations of the matric tensor $g\equiv \left\{ g_{\mu \nu
}\right\} $ and its quantum conjugate momentum operator.
\end{abstract}

\maketitle

\section{Introduction}

The principles of general covariance and of manifest covariance with respect
to the group of local point transformations (LPT-group, \cite{ein1,noi4}),
i.e., coordinate diffeomorphisms mutually mapping in each other different
general relativistic (GR) frames:\textbf{\ }%
\begin{equation}
r\Leftrightarrow r^{\prime }=f(r),  \label{LPT}
\end{equation}%
with $r\equiv \left\{ r^{\mu }\right\} $ and $r^{\prime }\equiv \left\{
r^{\prime \mu }\right\} $ denoting $4-$positions in the two frames, lie at
the foundation of all relativistic theories and of the related physical laws.

In fact, although the choice of special coordinate systems is always
legitimate for all\ physical systems, either classical or quantum, the
intrinsic objective nature of the physical laws that characterize them,
including possible symmetry transformations, makes them manifestly frame
independent. For the same reason, since LPTs preserve by construction the
differential-manifold structure of space-time, these principles represent
also a cornerstone of the so-called Standard Formulation of General
Relativity (SF-GR), namely the Einstein field equations and the
corresponding classical treatment of the gravitational field \cite%
{LL,gravi,wald}.

The same principles however, should apply to relativistic statistical\
mechanics and the very foundations of quantum field theory. The significance
of relativistic statistical mechanics is also of great importance in the
framework of general relativity and cosmology, including both classical and
quantum theories. This paper deals in particular with the problem of the
formulation of the theory of Quantum Gravity (QG). Despite major theoretical
developments achieved in the past, a theory of this type, i.e., fulfilling
the same principles, has remained until very recently \cite%
{noi1,noi2,noi5,noi6} largely unsolved. The fundamental reason is that - as
displayed in Ref.\cite{noi5} - a corresponding manifestly-covariant, and
possibly constraint-free, classical Hamiltonian theory of SF-GR is actually
required for the completion of such a task. On the other hand, in the\
previous\ mainstream literature typically only non-manifestly covariant
Hamiltonian theories of GR were developed. These are based on suitable
decompositions or foliations of space-time, i.e., the adoption of particular
subsets of GR-frames or coordinate systems and non-tensor
Lagrangian/Hamiltonian variables, which typically involve the singling out
of the coordinate time\ to prescribe the dynamical evolution of metric
tensor hypersurfaces (exhaustive developments of the issue can be found in
Refs.\cite{ADM,zzz2,alcu,Vaca2,Vaca5,Vaca6}).

Nevertheless, the mathematical framework to be adopted for the construction
of manifestly-covariant classical Lagrangian and Hamiltonian theories is
well-established both for particle dynamics \cite{EPJ1,EPJ2,EPJ3,EPJ4,EPJ5}
as well in continuum field theory, where it is known as the DeDonder-Weyl
formalism \cite{donder,weyl,sym3,sym4,sym5,sym6,sym7,sym8,sym9}. This type
of formulation has been developed and applied consistently to the case of
the gravitational field described by SF-GR only recently in Refs.\cite%
{noi5,noi6}, providing theories of covariant classical gravity (CCQ) and
covariant quantum gravity (CQG). In both cases the classical and quantum
Hamiltonian structures are realized in the framework of a so-called \emph{%
background space-time} picture, namely requiring that a prescribed
space-time $\left( \mathbf{Q}^{4},\widehat{g}\right) $ exists whose metric
tensor $\widehat{g}\equiv \left\{ \widehat{g}_{\mu \nu }(r)\right\} $ is
considered a prescribed classical field which determines the geometric
properties of the background space-time and, either classical or quantum,
variational tensor fields. More precisely, for this purpose $\left( \mathbf{Q%
}^{4},\widehat{g}\right) $\textbf{\ }is taken as a differentiable Lorentzian
manifold with signature $(+,-,-,-)$\ or analogous permutations, with $%
\mathbf{Q}^{4}$ denoting the $4-$dimensional Riemann space-time and $%
\widehat{g}_{\mu \nu }\left( r\right) $\ the metric tensor.

Nevertheless, historically basic conceptual problems which lie at the very
foundation of QG as a quantum theory of the gravitational field as described
classically by the Einstein equations, have been called into question. As
shown in Ref.\cite{noi5} a quantum theory of this type has conceptual
implications also for cosmology in connection with the possible existence of
massive gravitons associated with a non-vanishing cosmological constant $%
\Lambda $. Therefore, CQG-theory is expected to represent as well a
candidate quantum theory of the universe, i.e., a basis for Quantum
Cosmology. In view of these considerations, the issues addressed here are
the following ones:

\begin{itemize}
\item \emph{ISSUE \#1: the canonical} $g-$\emph{quantization -}\ The issue
concerns the quantization, here referred to as $g-$\emph{quantization,}%
\textbf{\ }of the classical canonical state $x=\left\{ g,\pi \right\} ,$\
with $\pi \equiv \left\{ \pi _{\mu \nu }\right\} $\ being the classical
reduced-dimensional canonical momentum conjugate to the continuum field
tensor $g\equiv \left\{ g_{\mu \nu }(r)\right\} $. This is realized by means
of a correspondence principle between the classical state $x=\left\{ g,\pi
\right\} $\ and the corresponding quantum variables $x^{(q)}=\left\{
g^{(q)}\equiv g,\pi ^{(q)}\right\} ,$ with $\pi ^{(q)}\equiv \left\{ \pi
_{\mu \nu }^{(q)}\right\} $\ being the corresponding quantum operator. In
such a context the question arises of\emph{\ }the \emph{unique prescription}
of the quantum-wave function and the corresponding quantum wave-equation
associated with $g-$\emph{quantization. }These should\ be\ understood
respectively as \emph{quantum wave-function} and \emph{quantum wave-equation
of the universe} and therefore to hold for arbitrary realizations of the
background space-time.\emph{\ }According to Ref.\cite{noi5} the $4-$scalar
(i.e., obtained by saturation of $4-$tensors) quantum state $\psi $ should
dynamically evolve with respect to an invariant proper-time parameter $s$\
defining the canonical Hamiltonian flow. Hence, besides $g\equiv \left\{
g_{\mu \nu }(r)\right\} ,$ $\psi $ is parametrized in terms of the
prescribed metric tensor $\widehat{g}(r)$ of the background space-time as
well as the $4-$position $r^{\mu }$ and the proper time $s$, whose physical
meaning in the context of QG remains nevertheless to be specified.

\item \emph{ISSUE \#2: the quantum Hamilton variational principle} - This is
about the search for a \emph{variational principle} for the same quantum
wave equation which may provide an additional "\textit{a posteriori}"
justification of its physical validity. The form of the same equation, in
fact, should be consistent with the existence of real symmetric functionals
which are bilinear with respect to the quantum wave function $\psi $, while
satisfying the principle of manifest-covariance for the Hamiltonian
functional. In agreement with the variational setting developed for the
classical derivation of the Einstein equations, also in the quantum regime
we seek the implementation of a \emph{synchronous variational principle}
characterized by having integral differential $4-$volume and/or line
elements which are left invariant during synchronous variations (see details
in Ref.\cite{noi1}).

\item \emph{ISSUE \#3: the }$g-$\emph{quantization Heisenberg inequalities -
}This concerns the\textbf{\ }problem of \emph{quantum measurement} and more
precisely the possible validity of a suitable \emph{Heisenberg principle}
which may provide inequalities appropriate for the treatment of\textbf{\ }$%
g- $\emph{quantization} and relate the standard deviations of the quantum
observables\textbf{\ }$x^{(q)}$.
\end{itemize}

The goal of this paper is to set these problems in the axiomatic framework
of CQG-theory earlier developed in Ref.\cite{noi5}. Such a theory, realizing
a manifestly-covariant canonical quantization of the classical Hamiltonian
state $x=\left\{ g,\pi \right\} ,$ is based on the assumption that the
corresponding quantum state represented by a single $4-$scalar wave-function
$\psi $ advances in proper-time by means of a non-stationary quantum-wave
equation (the CQG-wave equation), or equivalently of a corresponding set of
non-stationary quantum hydrodynamic equations.

More precisely, for this purpose, first, in reference to \emph{ISSUE \#1}
the problem is set in the context of the \emph{Hamilton-Jacobi (HJ) }$g-$%
\emph{quantization} formulated for the classical canonical state $x=\left\{
g,\pi \right\} $. As we intend to show, in such a context the CQG-wave
equation emerges uniquely from the classical Hamilton-Jacobi equation by
invoking quantization rules relating Poisson brackets to quantum commutators
and definition of conjugate quantum operators of field observables. The
resulting quantum wave equation does not require any independent
postulation, but rather is found to be naturally associated with the
classical Hamiltonian and Hamilton-Jacobi theories of GR. Second, in
reference to \emph{ISSUE \#2, }a suitable \emph{quantum Hamilton variational
principle} is established, which recovers the characteristic variational
property of quantum field theory. More precisely, the same quantum
wave-equation as well as the related quantum hydrodynamic equations are
shown to be variational both with respect to the quantum wave-function and
the corresponding quantum fluid fields. Finally, in reference to \emph{ISSUE
\#3}, a crucial property of quantum measurements performed in the context of
$g-$\emph{quantization} is displayed, lying in the validity of suitable
\emph{Heisenberg inequalities. }In particular we intend to show that, for
arbitrary quantum wave-functions which are solutions of the same CQG-wave
equation, the standard deviations associated with quantum measurements of $%
g_{\mu \nu }$ and $\pi _{\mu \nu }^{(q)}$ satisfy two different
inequalities, referred to respectively as \emph{first} and \emph{second
Heisenberg inequalities}.

In this connection, it is important to remark that one of the objectives of
CQG-theory is the search of solutions for the quantum-gravity wave-function $%
\psi $\ in contexts of physical and astrophysical interest, for example
those characterized by the occurrence of strong-gravity effects and where
quantum phenomena may play a relevant role. In particular, this generally
involves the determination of both proper-time dependent and stationary
solutions. In the second case the issue concerns the eigenvalue problems
associated with the stationary CQG-wave equation arising both in the case of
vacuum as well as in the presence of external sources. Thus, in particular,
the goal concerns the determination of the corresponding energy spectrum of
the quantum Hamiltonian operator and the conditions warranting its discrete
or continuous structure. Ultimately, a theory of this type should yield
information about phenomenological properties of gravitons and related
observational features, like mass, energy and quantum dynamics. An
application of this type of CQG-theory has been first proposed in Ref. \cite%
{noi6}, where the case of the stationary vacuum CQG-wave equation was
studied in a cosmological regime characterized by non-vanishing cosmological
constant $\Lambda $. A stationary equation for the CQG-state in terms of the
4-scalar invariant-energy eigenvalue was constructed for the corresponding
quantum Hamiltonian operator holding in the harmonic approximation, i.e.,
assuming to have oscillation of the quantum field $g_{\mu \nu }$\ suitably
close to the classical background tensor $\widehat{g}_{\mu \nu }$. The
conditions for the existence of a discrete invariant-energy spectrum were
determined, providing a possible estimate for the graviton mass in the
cosmological framework considered, together with the interpretation about
the quantum origin of the cosmological constant in terms of the graviton
Compton wavelength. The achievements of the present research are intended to
provide a theoretical framework that can be applied to further investigate
the phenomenology of such physical solutions. In particular, noting that the
predicted existence of gravitons by CQG-theory is directly related to the
quantum behavior of the field\ $g_{\mu \nu }$, it follows that the proof of
existence of Heisenberg inequalities for $g_{\mu \nu }$ and $\pi _{\mu \nu
}^{(q)}$\ are expected to constraint in turn also the observational
challenges of graviton particles themselves. Accordingly, the dynamics of
quantum gravitational fields and quantum gravitons are subject to
uncertainty effects which can be predicted on the basis of the theory
proposed below.

The scheme of presentation is as follows. First, in Section 2 the classical
Hamiltonian structure of GR is recalled, which is determined in the
framework of a manifestly-covariant approach. This leads to the introduction
of both Hamilton and Hamilton-Jacobi equations underlying the gravitational
field equations of GR. In Section 3 the Hamilton-Jacobi $g$-quantization
scheme is displayed, yielding a relationship between classical and quantum
Hamilton-Jacobi representations and yielding in this way a representation of
the quantum wave equation in manifestly-covariant form. Section 4 deals with
the formulation of a synchronous quantum Hamilton variational principle for
the quantum gravity wave equation. In Section 5 the quantum hydrodynamic
equations corresponding to the quantum wave equation are derived, upon
invoking the Madelung representation for the quantum wave function. The same
equations are then shown to be variational too, following from the same
synchronous variational principle leading to the CQG-wave equation. Then, in
Section 6 the validity is proved of suitable Heisenberg inequalities, while
concluding remarks are drawn in Section 7.

\section{The classical \textbf{Hamiltonian structure of SF-GR}}

In this section the Hamiltonian structure of space-time $\left\{ x=(g,\Pi
),H\right\} $ introduced in Ref.\cite{noi2} and the derivation of its
corresponding reduced-dimensional representation are recalled. This follows
by identifying the variational canonical variables $g$ and $\Pi $
respectively with the second-order real tensor fields $g\equiv \left( g_{\mu
\nu }(r)\right) $ and the third-order canonical momentum $\Pi \equiv \left(
\Pi _{\mu \nu }^{\alpha }(r)\right) $, and $H$ with a suitable variational
Hamiltonian density.\ As a consequence the resulting continuum Hamilton
equations (\emph{Extended Hamilton equations of SF-GR}) take the form%
\begin{equation}
\left\{
\begin{array}{c}
\nabla _{\alpha }g^{\mu \nu }=\frac{\partial H}{\partial \Pi _{\mu \nu
}^{\alpha }}, \\
\nabla _{\alpha }\Pi _{\mu \nu }^{\alpha }(r)=-\frac{\partial H}{\partial
g^{\mu \nu }},%
\end{array}%
\right.  \label{1}
\end{equation}%
with $\nabla _{\alpha }$ denoting hereon the covariant derivative in which
the standard connections (Christoffel symbols) are evaluated with respect to
the prescribed metric tensor $\widehat{g}(r)\equiv \left\{ \widehat{g}_{\mu
\nu }(r)\right\} $ of the background space-time $\left\{ \mathbf{Q}^{4},%
\widehat{g}(r)\right\} $ (see related discussion given in Ref.\cite{noi4}).
Here the Hamiltonian density $H\equiv H(x,\widehat{g}(r),r)$ and the related
effective kinetic ($T$) and potential ($V$) densities are taken respectively
of the form%
\begin{equation}
\left\{
\begin{array}{c}
\left. H\equiv T+V,\right. \\
\left. T\equiv \frac{1}{2\kappa f(h)}\Pi _{\mu \nu }^{\alpha }\Pi _{\alpha
}^{\mu \nu },\right. \\
\left. V\equiv \sigma V_{o}\left( g,\widehat{g}\right) +\sigma V_{F}\left( g,%
\widehat{g},r\right) ,\right.%
\end{array}%
\right.  \label{KINETIC ENERGY DENSITY}
\end{equation}%
with $\kappa $\ being the dimensional constant $\kappa =\frac{c^{3}}{16\pi G}
$\ and\ $G$\ being the universal gravitational constant. In addition, here $%
h $\ denotes the "variational" weight-factor first introduced in Ref.\cite%
{noi1}, which is also crucial for establishment of the same canonical
equations (\ref{1}), and is actually a suitably-prescribed function of the
variational tensor field $g(r)\equiv \left\{ g_{\mu \nu }(r)\right\} $,\
namely%
\begin{equation}
h=\left( 2-\frac{1}{4}g^{\alpha \beta }g_{\alpha \beta }\right) .
\label{variational weight-factor}
\end{equation}%
It is important to recall here that in the framework of the synchronous
classical variational principle, variational and prescribed metric tensors
are allowed to possess different properties and to satisfy distinctive
constraints. In particular, one has that $\widehat{g}^{\alpha \beta }%
\widehat{g}_{\alpha \beta }=\delta _{\alpha }^{\alpha }$, while at the same
time $g^{\alpha \beta }g_{\alpha \beta }\neq \delta _{\alpha }^{\alpha }$.
As a consequence, in general $h\left( g\right) \neq 1$, while it must be $%
h\left( \widehat{g}\right) =1$ identically. In addition, in Eq.(\ref{KINETIC
ENERGY DENSITY}) $f(h)$\ and $\sigma $\ are \emph{multiplicative gauge }%
functions which according to the Ref.\cite{noi5} should be respectively
identified with $f(h)=1$ and $\sigma =-1$. More precisely, according to the
notations reported in Ref.\cite{noi4}, the two $4-$scalars $V_{o}\left( g,%
\widehat{g}\right) $ and $V_{F}\left( g,\widehat{g},r\right) $ represent
respectively the vacuum and external-field contributions
\begin{equation}
\begin{array}{c}
\left. V_{o}\left( g,\widehat{g}\right) \equiv h\kappa \left[ g^{\mu \nu }%
\widehat{R}_{\mu \nu }-2\Lambda \right] ,\right. \\
\left. V_{F}\left( g,\widehat{x},r\right) \equiv hL_{F}\left( g,\widehat{x}%
,r\right) ,\right.%
\end{array}
\label{POT-ENERGY-SOURCES-2}
\end{equation}%
with $\widehat{R}_{\mu \nu }\equiv \left. R_{\mu \nu }(g(r))\right\vert
_{g(r)\equiv \widehat{g}(r)}$\ and $\Lambda $ denoting in $V_{o}$ the Ricci
tensor evaluated with respect to the prescribed metric tensor\ $\widehat{g}%
(r)$ and the cosmological constant. Accordingly, $L_{F}\left( g,\widehat{x}%
,r\right) $\ denotes a\ suitable $4-$scalar function depending on external
sources (for classical sources see the appropriate variational prescriptions
reported in Ref.\cite{noi1}).

Finally, for completeness, it is worth recalling that, as shown in Ref.\cite%
{noi2}, the extended Hamilton equations (\ref{1}) are variational, i.e.,
they coincide with the Euler-Lagrange equations of the synchronous classical
Hamilton variational principle%
\begin{equation}
\delta S(x,\widehat{x})=0,  \label{Hamilton principle}
\end{equation}%
with $S(x,\widehat{x})$\ denoting the classical Hamilton variational
functional%
\begin{eqnarray}
S(x,\widehat{x}) &=&\int_{\mathbf{Q}^{4}}d\Omega L(x,\widehat{g}(r),r), \\
L(x,\widehat{g}(r),r) &=&\Pi _{\mu \nu }^{\alpha }\nabla _{\alpha }g^{\mu
\nu }-H(x,\widehat{g}(r),r).
\end{eqnarray}

\subsection{\textbf{Reduced-dimensional Hamiltonian representation}}

The construction of the reduced dimensional representation for Eqs.(\ref{1})
is realized, first, by means of the \emph{splitting representation }for the
canonical momentum $\Pi _{\mu \nu }^{\alpha }(r)$, i.e., its representation
in terms of the direct product of two tensors, respectively first and
second-order tensors, of the form%
\begin{equation}
\Pi _{\mu \nu }^{\alpha }(r)=t^{\alpha }(r)\pi _{\mu \nu }(r).
\label{splitting}
\end{equation}%
Here $t^{\alpha }(r)$ denotes a real $4-$vector constructed to be a unit $4$%
-vector, i.e., such that%
\begin{equation}
\widehat{g}_{\alpha \nu }(r)t^{\alpha }(r)t^{\nu }(r)\equiv t^{\alpha
}(r)t_{\alpha }(r)=1,  \label{3}
\end{equation}%
and fulfilling identically the divergence-free condition%
\begin{equation}
\nabla _{\alpha }t^{\alpha }(r)=0.  \label{4}
\end{equation}%
The second step consists in the introduction of a proper-time
parametrization ($s-$parametrization) for the tensor fields, i.e., of the
form%
\begin{eqnarray}
&&\left. \left\{ g(s),\pi (s)\right\} =\left\{ g(r(s)),\pi (r(s))\right\}
,\right.  \label{S-PARAMETRIZATION} \\
&&\left. \widehat{g}(s)=\widehat{g}(r(s)),\right.
\end{eqnarray}%
in terms of suitable \emph{space-time curves} $\left\{ r(s),t(s),s\in
I\equiv
\mathbb{R}
\right\} $. The same curves belong to the background space-time $\left\{
\mathbf{Q}^{4},\widehat{g}(r)\right\} $, with $s$ denoting the proper time
defined on the same space-time for subluminal trajectories and $t(s)$
satisfying by construction for all $s\in I$ the constraints (\ref{3}) and (%
\ref{4}). A particular realization of such curves is provided by geodetics
of the metric $\widehat{g}\equiv \left\{ \widehat{g}_{\mu \nu }(r)\right\} $%
, namely curves fulfilling the initial-value problem%
\begin{equation}
\left\{
\begin{array}{c}
\frac{dr^{\mu }(s)}{ds}=t^{\mu }(s), \\
\frac{Dt^{\mu }(s)}{Ds}=0, \\
r^{\mu }(s_{o})=r_{o}^{\mu }, \\
t^{\mu }(s_{o})=t_{o}^{\mu },%
\end{array}%
\right.  \label{5}
\end{equation}%
where $\left( r_{o}^{\mu },t_{o}^{\mu }\right) $ represent respectively an
arbitrary $4-$position of $\left\{ \mathbf{Q}^{4},\widehat{g}(r)\right\} $
and an arbitrary tangent vector\ fulfilling Eq.(\ref{3}). Since at any point
of $\left( \mathbf{Q}^{4},\widehat{g}\right) $\ infinite geodetic curves
exist, this shows that the decomposition (\ref{splitting}), which fulfills
by construction the constraint equations (\ref{3}) and (\ref{4}), is always
possible. Thus in terms of the parametrization (\ref{S-PARAMETRIZATION}) and
thanks to Eqs.(\ref{splitting}), (\ref{3}) and (\ref{4}) the extended
canonical equations (\ref{1}) imply the \emph{reduced Hamilton equations of
SF-GR}, namely
\begin{equation}
\left\{
\begin{array}{c}
\frac{d}{ds}g^{\mu \nu }(s)=\frac{\partial H_{R}}{\partial \pi _{\mu \nu }(s)%
}, \\
\frac{d}{ds}\pi _{\mu \nu }(s)=-\frac{\partial H_{R}}{\partial g^{\mu \nu
}(s)},%
\end{array}%
\right.  \label{REDUCED HAM-EQS}
\end{equation}%
subject to the initial conditions
\begin{equation}
\left\{
\begin{array}{c}
g^{\mu \nu }(s_{o})=g_{(o)}^{\mu \nu }(r(s_{o})), \\
\pi _{\mu \nu }(s_{o})=\pi _{(o)\mu \nu }(r(s_{o})),%
\end{array}%
\right.
\end{equation}%
with $\left\{ g_{(o)}^{\mu \nu }(r(s_{o})),\pi _{(o)\mu \nu
}(r(s_{o}))\right\} $ denoting suitable initial tensor fields. Here, $\frac{d%
}{ds}$ denotes for greater generality the covariant\emph{\ }$s-$derivative%
\emph{, }i.e., the substantial derivative
\begin{equation}
\frac{d}{ds}=\left. \frac{\partial }{\partial s}\right\vert _{r}+\left.
t^{\alpha }\nabla _{\alpha }\right\vert _{s},  \label{s-derivative}
\end{equation}%
with $\left. \frac{\partial }{\partial s}\right\vert _{r}$ denoting the
partial $s-$derivative and $\left. t^{\alpha }\nabla _{\alpha }\right\vert
_{s}$ the convective derivative along a space-time trajectory $\left\{
r(s),t(s),s\in I\right\} .$ In addition,\ the partial derivatives $\frac{%
\partial H_{R}}{\partial \pi _{\mu \nu }(s)}$ and $\frac{\partial H_{R}}{%
\partial g^{\mu \nu }(s)}$ are performed with respect to the explicit
dependences only.\ As a consequence $H_{R}$ takes generally the form:%
\begin{equation}
\left\{
\begin{array}{c}
H=H_{R}\equiv T_{R}+V, \\
T_{R}(x,\widehat{g})\equiv \frac{1}{2kf(h)}\pi _{\mu \nu }\pi ^{\mu \nu },
\\
V(g,\widehat{g},r,s)\equiv \sigma V_{o}\left( g,\widehat{g},r\right) +\sigma
V_{F}\left( g,\widehat{g},r,s\right) ,%
\end{array}%
\right.
\end{equation}%
with $\sigma V_{o}$ and $\sigma V_{F}$ corresponding respectively to Eqs.(%
\ref{POT-ENERGY-SOURCES-2}) and where, in particular, for greater generality
$\sigma V_{F}$ is allowed to depend explicitly on $s$. As a consequence, in
the canonical equations (\ref{REDUCED HAM-EQS}):

a)\ the $s-$derivatives of the canonical state $x(s)=\left\{ g(s),\pi
(s)\right\} $\ must be intended of the type%
\begin{equation}
\frac{d}{ds}x(s)\equiv \left( \left. \frac{\partial }{\partial s}\right\vert
_{r}+\left. t^{\alpha }\nabla _{\alpha }\right\vert _{s}\right) x(s),
\label{covariant s-derivative of x}
\end{equation}%
i.e., to include also a possible explicit $s-$dependence;

b) the Hamiltonian density $H_{R}$\ by construction is actually taken in the
form\textbf{\ }%
\begin{equation}
H_{R}=H_{R}(x,\widehat{g}(r(s)),r(s),s;t),
\end{equation}%
with $t\equiv \left\{ t^{\mu }\right\} $ denoting a possible parametric
dependence on the tangent $4-$vector $t^{\mu }$ appearing through $r(s)$.
Such a dependence, arising because of Eq.(\ref{covariant s-derivative of x}%
), is nevertheless implicit and as such is not expected to affect the
solutions of the same reduced Hamilton equations.

Nevertheless, in case $H_{R}$\ does not depend explicitly on $r\equiv r(s)$%
,\ as may occur in the case of vacuum solutions and in the case $%
V_{o}=V_{o}\left( g,\widehat{g}\right) $, then necessarily $H_{R}$\ becomes
independent of $t^{\mu }$\ too. Notice, finally, that in order to apply
Hamilton-Jacobi quantization (see below), the dimensional-normalization%
\begin{equation}
\left\{
\begin{array}{c}
g^{\mu \nu }\rightarrow \overline{g}^{\mu \nu }=g^{\mu \nu } \\
\pi ^{\mu \nu }\rightarrow \overline{\pi }^{\mu \nu }=\frac{\alpha L}{k}\pi
^{\mu \nu } \\
H_{R}\rightarrow \overline{H}_{R}\equiv \overline{T}_{R}+\overline{V}=\frac{%
\alpha L}{k}H_{R} \\
\overline{V}(\overline{g},\widehat{g},r,s;t)\equiv \sigma \overline{V}%
_{o}\left( \overline{g},\widehat{g},r\right) +\sigma \overline{V}_{F}\left(
\overline{g},\widehat{g},r,s;t\right)%
\end{array}%
\right. ,  \label{CANONICAL-0}
\end{equation}%
can be conveniently introduced \cite{noi4}. Here $\left( \overline{g}^{\mu
\nu },\overline{\pi }^{\mu \nu }\right) $\ identifies the normalized
canonical state, with $L$\ and $\alpha $\ denoting a $4-$scalar scale-length
and an invariant parameter having the dimension of an action, both to be
suitably defined (see related discussion in Refs.\cite{noi4,noi5}). As a
result, the corresponding normalized Hamiltonian density and the effective
potentials set in dimensionally-normalized form, denoted by over-bars, are
given respectively by%
\begin{equation}
\left\{
\begin{array}{c}
\overline{H}_{R}\equiv \overline{T}_{R}+\overline{V} \\
\overline{T}_{R}(x,\widehat{g})\equiv \frac{1}{2\alpha Lf(h)}\overline{\pi }%
_{\mu \nu }\overline{\pi }^{\mu \nu } \\
\overline{V}(g,\widehat{g},r,s;t)\equiv \sigma \overline{V}_{o}\left( g,%
\widehat{g},r\right) +\sigma \overline{V}_{F}\left( g,\widehat{g}%
,r,s;t\right) \\
\overline{V}_{o}\left( \overline{g},\widehat{g},r\right) \equiv h\alpha
Lf\left( h\right) \left[ g^{\mu \nu }\widehat{R}_{\mu \nu }-2\Lambda \right]
\\
\overline{V}_{F}\left( g,\widehat{g},r,s;t\right) \equiv \frac{\alpha L}{k}%
V_{F}\left( g,\widehat{g},r,s;t\right) .%
\end{array}%
\right. .  \label{FINAL}
\end{equation}%
In the remainder, the constitutive equations (\ref{FINAL}) will be adopted
throughout the paper, dropping for simplicity the over-bar notation.

\subsection{Hamilton-Jacobi\textbf{\ equation}}

As shown in Ref.\cite{noi5} the reduced Hamilton equations (\ref{REDUCED
HAM-EQS}) of SF-GR are equivalent to the continuum Hamilton-Jacobi equation
for the Hamilton principal function $\mathcal{S\equiv S}\left( g(s),\widehat{%
g}(s),r(s),s;t\right) $ constructed in such a way that the canonical
momentum takes the form%
\begin{equation}
\pi ^{\mu \nu }=\frac{\partial \mathcal{S}\left( g(s),\widehat{g}%
(s),r(s),s;t\right) }{\partial g_{\mu \nu }(s)}.
\end{equation}%
This requires more precisely denoting $H_{R}\equiv H_{R}\left( g(s),\frac{%
\partial \mathcal{S}\left( g(s),\widehat{g}(s),r(s),s;\left[ t\right]
\right) }{\partial g},\widehat{g}(s),r(s),s;\left[ t\right] \right) $, with
the consequence that the Hamilton principal function $\mathcal{S}$ must be a
solution of the classical Hamilton-Jacobi equation:%
\begin{equation}
\frac{\partial \mathcal{S}}{\partial s}+H_{R}=0,  \label{HJ equation}
\end{equation}%
where $\frac{\partial }{\partial s}\equiv \frac{d}{ds}$ identifies the
substantial i.e., total covariant $s-$derivative defined by Eq.(\ref%
{s-derivative}) and the same $s-$derivative is performed keeping constant
both $g(s)$ and $\widehat{g}(s)$. In the particular case in which $r(s)$ and
$t$ are ignorable for the Hamiltonian density $H_{R}$ (i.e., as for vacuum
solutions), then also the Hamilton principal function becomes of the form $%
\mathcal{S}=\mathcal{S}\left( g(s),\widehat{g}(s),s\right) $.

\section{Hamilton-Jacobi $g$-quantization}

The key starting point for the quantization of the Hamiltonian system $%
\left\{ x,H_{R}\right\} $ developed by CQG-theory \cite{noi6} concerns the
notion of quantum state itself. This is to be intended as the quantum state
of a particle with spin-$2$\ which in the context of CQG is identified with
a massive graviton, i.e., having rest-mass $m_{o}>0$. In particular, this
refers to the assumption that such a state can be represented by a single
complex $4-$scalar quantum wave-function $\psi $. That such a prescription
is indeed possible in the context of a first-quantization approach adopted
by CQG-theory \cite{noi6}, having a classical background space-time $%
\widehat{g}_{\mu \nu }$ which is distinguished from the variational/quantum
field $g_{\mu \nu }$, follows from the fact that in such a case $\psi $\ can
always be identified with the tensor product of the form $\psi (s)=\widehat{g%
}_{\mu \nu }\psi ^{\mu \nu }(s)$. This establishes the relationship between
the $4-$scalar wave function $\psi (s)$ and the corresponding second-order $%
4-$tensor $\psi ^{\mu \nu }(s)$. In particular, one readily obtains that
necessarily $\psi ^{\mu \nu }(s)=\frac{1}{4}\psi (s)\widehat{g}^{\mu \nu }$,
so that in the present description the $4-$tensor $\psi ^{\mu \nu }(s)$ has
to be regarded as a derived quantity in terms of the wave-function $\psi (s)$
which is solved by the CQG-wave equation. Regarding the functional setting,
i.e., the prescription of the functional class of admissible wave-functions $%
\left\{ \psi \right\} $, it is assumed that $\psi $ are smoothly
differentiable complex functions which are parametrized in terms of the
space-time curves $\left\{ r(s),t(s),s\in I\equiv
\mathbb{R}
\right\} $ indicated above and fulfilling suitable boundary conditions for%
\textbf{\ }$s\rightarrow \pm \infty $\ and on the improper hyper-surface of
the configuration space $U_{g}$ (see below). In particular, $\psi (s)$ is
assumed to be a smooth function of the form $\psi (s)=\psi (g,\widehat{g}%
(s),r,s;t),$ with $g=\left\{ g_{\mu \nu }\right\} ,$ $s$ and $t=\left\{
t^{\mu }\right\} $ being independent variables which span respectively the
configuration space $U_{g},$ the time axis $I$ and the tangent space $%
TU_{g}, $ while
\begin{equation}
\rho (s)\equiv \left\vert \psi (s)\right\vert ^{2}  \label{quantum PDF}
\end{equation}%
identifies the corresponding quantum probability density function (\emph{%
quantum PDF}) on the configuration space. Here we shall assume for
definiteness that the variational tensors $g_{\mu \nu }$\ spanning $U_{g}$\
are symmetric so that $U_{g}$\ is identified with the subset of a linear
space $%
\mathbb{R}
^{10}$. In addition one can always require $g_{\mu \nu }$\ to be also
non--singular. This means, in other words, that for any unit $4-$vector $%
t^{\mu }$ fulfilling Eq.(\ref{3}) it must be that%
\begin{equation}
\Theta (\left\vert g_{\mu \nu }t^{\mu }t^{\nu }\right\vert ^{2})=1,
\label{constraint conditions g}
\end{equation}%
with $\Theta (x)=\left\{
\begin{array}{cc}
1 & (x>0) \\
0 & (x\leq 0)%
\end{array}%
\right. $\ denoting the strong Heaviside step function.\ In the context of $%
g-$quantization the configuration space is identified with $U_{g}$, so that
by construction its quantum probability is defined as%
\begin{equation}
P(U_{g})=\left\langle \psi |\psi \right\rangle \equiv
\int\limits_{U_{g}}d(g)\rho (s)=1,  \label{normalization}
\end{equation}%
with $d(g)\equiv \prod\limits_{\mu ,\nu =1,4}$ $d\widehat{g}_{\mu \nu }$\
denoting the canonical measure on $U_{g}$. In addition, both $r$ and $%
\widehat{g}(r)$ are considered functions of $s$, i.e., evaluated along the
classical space-time curves $\left\{ r(s),t(s),s\in I\equiv
\mathbb{R}
\right\} ,$ so that $r\equiv r(s)=\left\{ r^{\mu }(s)\right\} $ and $%
\widehat{g}(s)\equiv \widehat{g}(r(s))$. Finally, the wave -functions $\psi
(s)$ span by assumption a Hilbert space $\Gamma _{\psi },$ which is
finite-dimensional in the sense that it is defined on a continuum
configuration space $U_{g}$ having a finite dimension. This is endowed with
the scalar product%
\begin{equation}
\left\langle \psi _{a}|\psi _{b}\right\rangle \equiv
\int\limits_{U_{g}}d(g)\psi _{a}^{\ast }(g,\widehat{g}(r),r(s),s)\psi _{b}(g,%
\widehat{g}(r),r(s),s),  \label{Axiom 1-2}
\end{equation}%
with $\psi _{a,b}(s)\equiv \psi _{a,b}(g,\widehat{g}(r),r(s),s)$\ being
arbitrary elements of the Hilbert space $\Gamma _{\psi }$, where as usual $%
\psi _{a}^{\ast }$\ denotes the complex conjugate of $\psi _{a}$.

Based on these premises, the formal construction of CQG-theory is then based
on the adoption of two distinctive axioms related respectively to the
following two prescriptions:

\begin{itemize}
\item First, the canonical quantization rule - hereon referred as $g-$\emph{%
quantization rule} - prescribing the mapping between the classical and
quantum Hamiltonian structures%
\begin{equation}
\left( x=\left\{ g,\pi \right\} ,H_{R}\right) \Rightarrow \left(
x^{(q)}=\left\{ g^{(q)},\pi ^{(q)}\right\} ,H_{R}^{(q)}\right) .
\label{mapping-1}
\end{equation}%
This mapping is realized by the CQG-correspondence principle, namely%
\begin{equation}
\left\{
\begin{array}{c}
g_{\mu \nu }\rightarrow g_{\mu \nu }^{(q)}\equiv g_{\mu \nu } \\
\pi _{\mu \nu }\rightarrow \pi _{\mu \nu }^{(q)}\equiv -i\hslash \frac{%
\partial }{\partial g^{\mu \nu }} \\
H_{R}\rightarrow H_{R}^{(q)}=\frac{1}{f(h)}T_{R}^{(q)}(\pi ,\widehat{g})+V%
\end{array}%
.\right.  \label{map-3}
\end{equation}

\item Second, the quantum-wave equation advancing in proper-time the same
quantum state. This is provided by the CQG-wave equation\emph{\ }%
\begin{equation}
i\hslash \frac{\partial }{\partial s}\psi (s)=\left[ H_{R}^{(q)},\psi (s)%
\right] \equiv H_{R}^{(q)}\psi (s),  \label{QG-WAVW EQUATION}
\end{equation}%
with $\frac{\partial }{\partial s}$ denoting again the total covariant $s-$%
derivative defined by Eq.(\ref{s-derivative}) and $\left[ A,B\right] $\emph{%
\ }being the quantum commutator\emph{\ }$\left[ A,B\right] \equiv AB-BA$.
\end{itemize}

The goal of this section is to set the CQG-theory in the context of a
Hamilton-Jacobi quantization scheme (see for example Ref.\cite{EPJP-2015})
which permits us to determine immediately the precise form of the resulting
quantum wave equation displayed in Eq.(\ref{QG-WAVW EQUATION}). The starting
point is the prescription of the relevant classical canonical momenta in the
context of the Hamilton-Jacobi approach. By direct inspection of the
Hamilton-Jacobi equation recalled above in Eq.(\ref{HJ equation}), it
follows that \emph{\ }$g-$\emph{quantization} can be achieved in terms of
the two classical canonical momenta $\frac{\partial \mathcal{S}\left( g,%
\widehat{g},(r),s;t\right) }{\partial g_{\mu \nu }}$ and $\frac{\partial
\mathcal{S}\left( g,\widehat{g},(r),s;t\right) }{\partial s}$, identifying
respectively $\pi _{\mu \nu }$ and the canonical momentum conjugate to the
proper time $s$. In the context of the Hamilton-Jacobi quantization, the
appropriate correspondence principle for $g-$\emph{quantization,}\
establishing the relationship between the classical and quantum momenta and
Hamiltonian functions, is then provided by the mapping \textbf{\ }%
\begin{eqnarray}
&&\left. \pi _{\mu \nu }\equiv \frac{\partial \mathcal{S}\left( g,\widehat{g}%
,(s),r,s;\left[ t\right] \right) }{\partial g^{\mu \nu }}\rightarrow \pi
_{\mu \nu }^{(q)}\equiv -i\hbar \frac{\partial }{\partial g^{\mu \nu }}%
,\right.  \label{MAP-1} \\
&&\left. p\equiv -\frac{\partial \mathcal{S}\left( g,\widehat{g}(s),r,s;%
\left[ t\right] \right) }{\partial s}\rightarrow p^{(q)}\equiv -i\hbar \frac{%
\partial }{\partial s},\right.  \label{MPA-2} \\
&&\left. H_{R}\left( g,\frac{\partial \mathcal{S}\left( g,\widehat{g}(s),r,s;%
\left[ t\right] \right) }{\partial g},\widehat{g}(s),r,s;\left[ t\right]
\right) \rightarrow H_{R}^{(q)}.\right.  \label{MAP-3}
\end{eqnarray}%
with $\pi _{\mu \nu }^{(q)},$ $p^{(q)}$ and $H_{R}^{(q)}$ identifying the
quantum canonical momenta conjugate to $g_{\mu \nu }$ and $s$ respectively
and the quantum Hamiltonian operator. This will be referred to as \emph{%
Hamilton-Jacobi }$g-$\emph{quantization.} Notice that, in the same mapping
the "coordinates" $g\equiv \left\{ g_{\mu \nu }\right\} $\ and $s$\ remain
unchanged, i.e., so that they still formally coincide with the classical
ones. The mapping realized by Eqs.(\ref{MAP-1}), (\ref{MPA-2}) and (\ref%
{MAP-3}) implies the simultaneous validity of the two fundamental commutator
relations
\begin{equation}
\left[ \pi ^{(q)\alpha \beta },g_{\mu \nu }\right] =-i\hslash \delta _{\mu
}^{\alpha }\delta _{\nu }^{\beta },  \label{FIRST COMMUTATOR}
\end{equation}%
\begin{equation}
\left[ p^{(q)},s\right] =-i\hslash ,
\end{equation}%
together with%
\begin{equation}
\left[ g^{\alpha \beta },g_{\mu \nu }\right] =\left[ \pi ^{(q)\alpha \beta
},\pi _{\mu \nu }^{(q)}\right] =0.
\end{equation}%
Notice that since both $\pi ^{(q)\alpha \beta }$\ and $g_{\mu \nu }^{(q)}$\
are symmetric, Eq.(\ref{FIRST COMMUTATOR}) holds for arbitrary permutations
of the indexes. As a consequence, based on the classical Hamilton-Jacobi
equation (\ref{HJ equation}) the additional mapping
\begin{equation}
\fbox{$\frac{\partial \mathcal{S}}{\partial s}+H_{R}=0$}\Rightarrow \fbox{$%
\left\{ p^{(q)}+H_{R}^{(q)}\right\} \psi \left( s\right) =0$}
\end{equation}%
follows which implies validity of the quantum-wave equation indicated above
(see Eq.(\ref{QG-WAVW EQUATION})).

We stress that the same CQG-wave equation exhibits a number of distinctive
properties:

\begin{enumerate}
\item It is manifestly covariant, i.e., it retains its form under the action
of arbitrary local point transformations (\ref{LPT}), which preserve by
construction the differential manifold of the space-time $\left( \mathbf{Q}%
^{4},\widehat{g}(r)\right) $.

\item It is an evolution equation which is parametrized in terms of the
proper-time $s$, i.e., the Riemann distance which is associated with the
background space-time for subluminal trajectories.

\item It advances in proper-time the $4-$scalar wave-function $\psi (s)$,
the associated configuration-space quantum PDF being prescribed by Eq.(\ref%
{quantum PDF}).

\item The same wave equation holds in principle for arbitrary initial
conditions as well as for arbitrary external sources, as is appropriate for
the treatment of problems of QG and quantum cosmology.
\end{enumerate}

In particular, an interesting comparison can be made with the Wheeler-DeWitt
wave equation \cite{dew,Wheeler}. Indeed it is well-known that\ the
Wheeler-DeWitt wave equation realizes a Schr\"{o}dinger-like evolution
equation advancing in time the dynamics of the so-called "wave function of
the universe" \cite{hARTE-hAWKING}. Despite its formal analogy, however,
basic differences emerge, the most important one being that the
Wheeler-DeWitt wave equation is not manifestly covariant. The reason, as
earlier discussed (see Ref.\cite{noi6}), is that its time evolution is
parametrized with respect to the coordinate-time $t,$ which is not an
invariant $4-$scalar.

\section{The quantum Hamilton variational principle}

The goal of this section is to prove that the CQG-wave equation (\ref%
{QG-WAVW EQUATION}) admits also a variational formulation in terms of a
suitably-prescribed synchronous variational principle. In this regard a
basic prerequisite is that the variational principle yielding this equation
should be manifestly covariant. In other words the corresponding variational
functional, the appropriate class of variations as well as the resulting
Euler-Lagrange equations should all be covariant with respect to the
LPT-group.\ As we intend to show here such a requirement is non-trivial and
restricts the class of admissible quantum-wave equations for quantum gravity.

A key feature of the CQG-wave equation is that it is a hyperbolic evolution
equation which prescribes the dependence of the quantum state $\psi =\psi
(s) $ in terms of the proper-time $s$ which is associated with the
background curved space-time $\left( \mathbf{Q}^{4},\widehat{g}_{\mu \nu
}(r)\right) $ and parametrizes the space-time curves $\left\{ r(s),t(s),s\in
I\equiv
\mathbb{R}
\right\} $. Such an equation is a first order partial differential equation
with respect to $s$, to be supplemented by suitable initial conditions,
namely prescribing for all $r(s_{o})=r_{o}\in \left( \mathbf{Q}^{4},\widehat{%
g}(r)\right) $ the condition $\psi (s_{o})=\psi _{o}(g,\widehat{g}%
(r_{o}),r_{o};t)$, as well as boundary conditions at infinity on the
improper boundary of configuration space $U_{g}$, i.e., letting $%
\lim_{g\rightarrow \infty }\psi (g,\widehat{g}(s),r(s),s;t)=0$.

Let us now show that the form of Eq.(\ref{QG-WAVW EQUATION}) indicated above
warrants that it is also variational, i.e., that in analogy with
quantum-wave equations known in quantum mechanics, it admits a
manifestly-covariant variational formulation. For the construction of the
variational principle one can indeed adopt a method analogous to that
followed in Ref.\cite{EPJP-2015} in the case\ of relativistic quantum
mechanics. The prerequisites are set by the following prescriptions:

\noindent \emph{Prescription \#1 -} A scalar product is defined on a
suitable extended configuration space, identified with the direct product $%
U_{g}\times I_{(s_{o},s_{1})}$,\ $U_{g}$\ being the configuration space
defined above and $I_{(s_{o},s_{1})}$\ the subset of the real axis $I\equiv
\mathbb{R}
,$\ namely the interval $I_{(s_{o},s_{1})}=\left[ s_{o},s_{1}\right] $.

\noindent \emph{Prescription \#2 -}\textbf{\ }A real functional $Q\left(
\psi ,\psi ^{\ast }\right) $\ of the variational wave function $\psi \equiv
\psi (s)$\ is defined which is symmetric with respect to the same scalar
product on the set $U_{g}$. The same function $\psi $\ is required to belong
to the functional class $\left\{ \psi (s)\right\} $\ of $C^{(2)}$
(continuous twice-differentiable) complex functions prescribed so that the
associated quantum PDF $\rho (s)$\ (see Eq.(\ref{quantum PDF})) fulfills on $%
U_{g}$\ the normalization condition (\ref{normalization}).

\noindent \emph{Prescription \#3 - }The same real functional and the
corresponding Lagrangian density ($L_{Q}$) are all $4-$scalars with respect
to generic background space-time $\left( \mathbf{Q}^{4},\widehat{g}%
(r)\right) .$

\noindent \emph{Prescription \#4 - }The Lagrangian density can be
represented in terms of the first-order differential operators associated
with the quantum canonical momenta, respectively $\pi _{\mu \nu }^{(q)}$ and
$p^{(q)}$ (see Eqs.(\ref{MAP-1}) and (\ref{MPA-2})) or their complex
conjugates $\pi _{\mu \nu }^{(q)\ast }$ and $p^{(q)\ast }$, acting
respectively on $\psi $ and its complex conjugate wave-function $\psi ^{\ast
}$.

In the present framework the variational functional\ corresponding to these
prescriptions is provided by a real symmetric $4-$scalar Hamilton functional
of the form%
\begin{equation}
Q\left( \psi ,\psi ^{\ast }\right)
=\int_{s_{o}}^{s_{1}}ds\int_{U_{g}}d(g)L_{Q}\left( \mathbf{w},\mathbf{w}%
^{\ast }\right) ,  \label{HAMILTON FUNCTIONAL}
\end{equation}%
where $s_{o},s_{1}\in I$\ are arbitrary proper times such that $s_{o}<s_{1}$%
\ (so that in particular a possible choice is realized by setting $%
s_{o}=-\infty ,$\ $s_{1}=+\infty $), while $d(g)$ represents the $4-$scalar
volume element of configuration space $U_{g}$. Furthermore, denoting $\psi
^{\ast }\equiv \psi ^{\ast }(s)$ the complex conjugate of $\psi \equiv \psi
(s)$ and $\mathbf{w}\equiv \left( \psi ,\frac{\partial \psi }{\partial
g_{\mu \nu }},\frac{\partial \psi }{\partial s}\right) $, then $\mathbf{w}%
^{\ast }$ is its complex conjugate. In particular, for consistency with
Prescriptions \#2-\#4, $L_{Q}\left( \mathbf{w},\mathbf{w}^{\ast }\right) $
must be identified with the $4-$scalar real Lagrangian density%
\begin{equation}
L_{Q}\left( \mathbf{w},\mathbf{w}^{\ast }\right) =i\hbar \frac{1}{2}\left[
\psi \frac{\partial }{\partial s}\psi ^{\ast }-\psi ^{\ast }\frac{\partial }{%
\partial s}\psi \right] +\frac{1}{2\alpha L}\left( i\hbar \frac{\partial }{%
\partial g_{\mu \nu }}\right) \psi ^{\ast }\left( -i\hbar \frac{\partial }{%
\partial g^{\mu \nu }}\right) \psi +V\psi ^{\ast }\psi .
\end{equation}%
In addition, consistent with Prescription \#1, the integral occurring in the
Hamilton functional $Q\left( \psi ,\psi ^{\ast }\right) $\ is performed
respectively: a) with respect to $s$,\ with integration carried out both
with respect to the explicit and implicit dependences contained in $\psi (g,%
\widehat{g}(s),r(s),s;t)$, and\ b) with respect to the explicit dependence
in terms of $g\equiv \left\{ g_{\mu \nu }\right\} $. This means that $%
Q\left( \psi ,\psi ^{\ast }\right) $\ can be equivalently represented as%
\textbf{\ }%
\begin{equation}
Q\left( \psi ,\psi ^{\ast }\right) =\int_{s_{o}}^{s_{1}}ds\left\langle
\left. \psi \right\vert K_{Q}\psi \right\rangle ,
\end{equation}%
with $\left\langle \left. \psi \right\vert K_{Q}\psi \right\rangle $\
representing the quantum expectation value (see Ref.\cite{noi5}) and $K_{Q}$%
\ being here the quantum operator%
\begin{eqnarray}
K_{Q} &=&K_{Q}^{(1)}+\frac{1}{2\alpha L}\left( i\hbar \overset{%
\Longleftarrow }{\frac{\partial }{\partial g^{\mu \nu }}}\right) \left(
-i\hbar \overset{\Longrightarrow }{\frac{\partial }{\partial g^{\mu \nu }}}%
\right) +V, \\
K_{Q}^{(1)} &=&i\hbar \frac{1}{2}\left[ \overset{\Longleftarrow }{\frac{%
\partial }{\partial s}}-\overset{\Longrightarrow }{\frac{\partial }{\partial
s}}\right] ,
\end{eqnarray}%
and where $\left( \overset{\Longleftarrow }{\frac{\partial }{\partial s}},%
\overset{\Longleftarrow }{\frac{\partial }{\partial g_{\mu \nu }}}\right) $\
and $\left( \overset{\Longrightarrow }{\frac{\partial }{\partial s}},\overset%
{\Longrightarrow }{\frac{\partial }{\partial g^{\mu \nu }}}\right) $\
identify the "bra" and\ "ket"\textbf{\ }operators acting on $\left\langle
\psi \right\vert \equiv \psi ^{\ast }$\ and $\psi \equiv \left\vert \psi
\right\rangle $\ respectively. Hence, $Q\left( \psi ,\psi ^{\ast }\right) $\
can be equivalently represented in terms of the quantum Hamiltonian operator
$H_{R}^{(q)}$\ and the scalar product defined for the waves functions
belonging to $\left\{ \psi \right\} $, namely it takes the form%
\begin{equation}
Q\left( \psi ,\psi ^{\ast }\right) =\int_{s_{o}}^{s_{1}}ds\left\langle
\left. \psi \right\vert \left[ -i\hbar \frac{\partial }{\partial s}%
+H_{R}^{(q)}\right] \psi \right\rangle ,  \label{HAMILTON FUNCTIONAL-2}
\end{equation}%
with these operators $-i\hbar \frac{\partial }{\partial s}$\ and\ $%
H_{R}^{(q)}$\ to be intended as ket operators acting on\ $\psi \equiv
\left\vert \psi \right\rangle $.

As a basic consequence the same operators, and hence also $K_{Q},$\ are all
symmetric while all the involved quantum operators have a $4-$tensor nature.
This means that the Hamilton functional (\ref{HAMILTON FUNCTIONAL}) (or
equivalently Eq.(\ref{HAMILTON FUNCTIONAL-2})) is such that: 1) $Q\left(
\psi ,\psi ^{\ast }\right) $\ is a real $4-$scalar functional, with all the
quantum operators and variables being represented by $4-$tensor quantities;
2) the quantum operators $i\hbar \frac{\partial }{\partial s}$\ and\ $%
H_{R}^{(q)}$, and hence also $K_{Q},$\ are all symmetric with respect to the
same scalar product.

Thus, introducing the functional class of variations $\left\{ \psi
_{1}\equiv \psi +\delta \psi \right\} $ it is assumed that all functions $%
\psi _{1}$ belong to the class of admissible wave-functions $\left\{ \psi
\right\} $ indicated above, while the variation $\delta \psi \equiv \delta
\psi (g,\widehat{g}(s),r(s),s;t)$\ is considered as a $4-$scalars complex
function. In other words, the same quantum wave equation (\ref{QG-WAVW
EQUATION}) must be uniquely determined by means of the quantum Hamilton
variational principle%
\begin{equation}
\delta Q\left( \psi ,\psi ^{\ast }\right) =0,  \label{HAMILT-VARIATIONAL}
\end{equation}%
to hold in a suitable class of variations in which the variations $\delta
\psi $\ (and similarly $\delta \psi ^{\ast }$) are considered as independent
$4-$scalar functions, with $\delta Q\left( \psi ,\psi ^{\ast }\right) $
denoting the Frechet derivative evaluated with respect to the same
variations, namely letting%
\begin{equation}
\delta Q\left( \psi ,\psi ^{\ast }\right) =\lim_{\alpha \rightarrow 0}\frac{%
Q\left( \psi +\alpha \delta \psi ,\psi ^{\ast }+\alpha \delta \psi ^{\ast
}\right) -Q\left( \psi ,\psi ^{\ast }\right) }{\alpha }.  \label{Frechet}
\end{equation}%
The same equation is required to realize a so-called synchronous variational
principle, i.e., performed again in terms of a synchronous variation
operator $\delta $\ \cite{noi1} which in this case leaves unchanged the line
element $ds$. This implies that the equation%
\begin{equation}
\delta \left( ds\right) =0  \label{constraint}
\end{equation}%
must hold identically. Such a differential constraint can be fulfilled by
suitable prescription of the variations $\delta \psi $\ and $\delta \psi
^{\ast }$ and in particular requiring that the parameter $\alpha $ in the
Frechet derivative (\ref{Frechet}) is independent of $s$. Moreover, we shall
assume that the same variations ($\delta \psi $\ and $\delta \psi ^{\ast }$%
)\ are considered independent and vanish on the boundary of $Q^{4}$\ and at
infinity for $s\rightarrow \pm \infty $. Then, by elementary algebra it
follows that the wave function $\psi $\ which is extremal for the functional
$Q\left( \psi ,\psi ^{\ast }\right) $\ must fulfill the Euler-Lagrange
equations%
\begin{eqnarray}
\frac{\delta Q\left( \psi ,\psi ^{\ast }\right) }{\delta \psi ^{\ast }} &=&0,
\label{E_L-1} \\
\frac{\delta Q\left( \psi ,\psi ^{\ast }\right) }{\delta \psi } &=&0.
\label{E-L-2}
\end{eqnarray}%
This proves at the same time the variational and manifest-covariant
properties of the quantum-wave equation (\ref{QG-WAVW EQUATION}). In fact,
thanks to the symmetry property of the Hamilton functional $Q\left( \psi
,\psi ^{\ast }\right) $,\ Eqs.(\ref{E_L-1}) and (\ref{E-L-2}) coincide
respectively with Eq.(\ref{QG-WAVW EQUATION}) and its complex conjugate, so
that the variational character of equation (\ref{QG-WAVW EQUATION}) is
established. Furthermore, thanks to the $4-$tensor property of the same
functional $Q\left( \psi ,\psi ^{\ast }\right) $\ and of the quantum
variables and operators,\ the same Euler-Lagrange equations (\ref{E_L-1})
and (\ref{E-L-2}) are all manifestly covariant.

These properties are distinctive of CQG-theory which depart from previous QG
literature. For example, let us consider again the comparison with the
Wheeler-DeWitt wave equation \cite{dew,Wheeler}. The latter equation is
indeed variational. Nevertheless, despite that fact that the same
variational principle is in some sense analogous to (\ref{HAMILT-VARIATIONAL}%
), it is also not manifestly-covariant. This arises, again, because the
Wheeler-DeWitt wave equation itself is based on the so-called $3+1$\
foliation of space-time, so that the variational principle occurring for the
same equation is not manifestly covariant. This problem however does not
arise in the framework of CQG-theory considered here.

\section{The variational quantum hydrodynamic equations}

A pre-requisite for establishing Heisenberg inequalities in the framework of
CQG-theory is the introduction of the concept of a quantum probability
density function and the derivation of the corresponding set of quantum
hydrodynamic equations (CQG-QHE) implied by the\ CQG-wave equation, to be
prescribed in conservative form in order to warrant conservation of quantum
probability.

In the present setting, the quantum probability density function (PDF)
associated with the CQG-state is identified with the real function $\rho
(s)\equiv \rho (g,\widehat{g}(r),r(s),s)$ and is prescribed as\emph{\ }%
\begin{equation}
\rho (s)\equiv \left\vert \psi (s)\right\vert ^{2},  \label{roro}
\end{equation}%
in formal analogy with the customary definition of quantum PDF holding in
non-relativistic quantum mechanics. Here $\rho (s)$ is a $4-$scalar and
represents the probability density of the Lagrangian field variable $g\equiv
\left\{ g_{\mu \nu }\right\} $\ in the volume element $d(g)$\ belonging to
the configuration space $U_{g}.$\emph{\ }By assumption the probability $P(A)$
of an arbitrary subset $A\subseteq U_{g}$ is normalized, in the sense that
for arbitrary $(\widehat{g}(r),r(s),s)$ it must be%
\begin{equation}
P(A)\equiv \left\langle \psi |\delta _{A}(g)\psi \right\rangle \equiv
\int\limits_{U_{g}}d(g)\rho (s)\delta _{A}(g),  \label{normaliz}
\end{equation}%
with $\delta _{A}(g)$ being the characteristic function of the set $A$. In a
similar way, given the validity of Eq.(\ref{roro}), one can define the real
function $S^{(q)}(s)\equiv S^{(q)}(g,\widehat{g}(r),r(s),s)$\emph{\ }as%
\begin{equation}
S^{(q)}(s)\equiv \hslash \arcsin \text{h }\left\{ \frac{\psi (s)-\psi ^{\ast
}(s)}{2\sqrt{\rho (s)}}\right\} ,  \label{esse}
\end{equation}%
which represents, on the configuration space $U_{g}$, the quantum
phase-function associated with the same CQG-state $\psi (s)$. Hence, in
terms of the real $4-$scalar field functions $\rho (s)$\ and $S^{(q)}(s)$
prescribed respectively by Eqs.(\ref{roro}) and (\ref{esse}), the CQG-state
defined by the complex function $\psi (s)$ can be cast in the equivalent
form of the Madelung exponential representation
\begin{equation}
\psi (s)=\sqrt{\rho (s)}\exp \left\{ \frac{i}{\hslash }S^{(q)}(s)\right\} ,
\label{Madelung}
\end{equation}%
i.e., the quantum fluid fields $\left\{ \rho (s),S^{(q)}(s)\right\} $
identifying respectively the quantum PDF and the quantum phase-function.
Once the Madelung representation is invoked, it is possible to replace the
single CQG-wave equation (\ref{QG-WAVW EQUATION}) for the complex
wave-function $\psi (s)$ with the equivalent set of CQG-QHE realized
respectively by a continuity equation for the real quantum PDF $\rho (s)$
and a quantum Hamilton-Jacobi equation for the real quantum phase-function $%
S^{(q)}(s)$.

In particular, invoking the Madelung representation for $\psi (s)$, given
the definition of the Hamiltonian operator $H_{R}^{(q)}$ according to Eq.(%
\ref{map-3}) and imposing the constraint condition $f(h)=1$, the following
set of real PDEs are obtained from the CQG-wave equation:%
\begin{eqnarray}
\frac{\partial \rho (s)}{\partial s}+\frac{\partial }{\partial g_{\mu \nu }}%
\left( \rho (s)V_{\mu \nu }(s)\right) &=&0,  \label{cont1} \\
\frac{\partial S^{(q)}(s)}{\partial s}+H_{c}^{(q)} &=&0.  \label{hj-2}
\end{eqnarray}%
These are referred to as\ CQG-quantum continuity equation and CQG-quantum
Hamilton-Jacobi equation advancing in proper-time respectively $\rho (s)$\
and $S^{(q)}(s).$ Here the quantum hydrodynamics fields $\rho (s)\equiv \rho
(g,\widehat{g},s)$\ and $S^{(q)}(s)\equiv S^{(q)}(g,\widehat{g},s)$\ are\
assumed to depend smoothly on the tensor field $g\equiv \left\{ g_{\mu \nu
}\right\} $\ spanning the configuration space $U_{g}$\ and in addition to
admit a Lagrangian path-parametrization in terms of the\ geodetics $%
r(s)\equiv \left\{ r^{\mu }(s)\right\} $\ associated locally with the
prescribed field $\widehat{g}(r)\equiv \left\{ \widehat{g}_{\mu \nu
}(r)\right\} $. The $4-$tensor $V_{\mu \nu }(s)$\ is prescribed instead as
\begin{equation}
V_{\mu \nu }(s)=\frac{1}{\alpha L}\frac{\partial S^{(q)}}{\partial g^{\mu
\nu }}.  \label{F-EQ-1}
\end{equation}%
Finally, $H_{c}^{(q)}$\ identifies the effective quantum Hamiltonian density%
\begin{equation}
H_{c}^{(q)}=\frac{1}{2\alpha L}\frac{\partial S^{(q)}}{\partial g^{\mu \nu }}%
\frac{\partial S^{(q)}}{\partial g_{\mu \nu }}+V_{QM}+V,  \label{F-EQ-2}
\end{equation}%
with $V\equiv V(g,\widehat{g}(r),r,s)$\ being the effective potential
density and\ $V_{QM}$\ a potential density denoted as Bohm-like effective
quantum potential which is prescribed as%
\begin{equation}
V_{QM}(g,\widehat{g}(r),r,s)=\frac{\hslash ^{2}}{8\alpha L}\frac{\partial
\ln \rho }{\partial g^{\mu \nu }}\frac{\partial \ln \rho }{\partial g_{\mu
\nu }}-\frac{\hslash ^{2}}{4\alpha L}\frac{\partial ^{2}\rho }{\rho \partial
g_{\mu \nu }\partial g^{\mu \nu }}.
\end{equation}%
The effective potential $V_{QM}$ is analogous to the well-known Bohm
potential met in non-relativistic quantum mechanics (see for example Refs.%
\cite{FoP1,FoP2}), its physical origin arising due to the non-uniformity of
the quantum PDF $\rho $, namely such that generally it must be $\frac{%
\partial }{\partial g_{\mu \nu }}\rho (s)\neq 0$. Finally, it must be
stressed that, as discussed in Ref.\cite{noi6}, the constraint condition $%
f(h)=1$ is required in order to warrant the quantum unitarity principle,
namely the conservation of quantum probability, thus resolving at quantum
level the indeterminacy on the prescription of the function $f\left(
h\right) $ arising in the classical Hamilton-Jacobi theory of GR\ (see also
Ref.\cite{noi5}).

A theoretical implication of the validity of the variational principle (\ref%
{HAMILT-VARIATIONAL}) can be obtained if in the functional (\ref{HAMILTON
FUNCTIONAL-2}) the wave functions $\psi $ and $\psi ^{\ast }$ are
represented in terms of the Madelung representation (\ref{Madelung}). In
fact, in terms of the same fields the Hamilton functional (\ref{HAMILTON
FUNCTIONAL-2}) then becomes%
\begin{equation}
Q\left( \psi ,\psi ^{\ast }\right)
=\int_{s_{o}}^{s_{1}}ds\int_{U_{g}}d(g)\left\{ \frac{i\hbar }{2}\left[ \frac{%
\partial \rho (s)}{\partial s}+\frac{\partial }{\partial g_{\mu \nu }}\left(
\rho (s)V_{\mu \nu }(s)\right) \right] +\rho (s)\left[ \frac{\partial
S^{(q)}(s)}{\partial s}+H_{c}^{(q)}\right] \right\} ,  \label{LAST-ONE}
\end{equation}%
with $V_{\mu \nu }(s)$\ and $H_{c}^{(q)}$\ being respectively the
second-order $4-$tensor and the effective quantum Hamiltonian defined by
Eqs.(\ref{F-EQ-1}) and (\ref{F-EQ-2}). For definiteness, let us require,
consistent with Prescription \#2,\ that the normalization (\ref%
{normalization}) applies. As a consequence, the functional class $\left\{
\psi (s)\right\} $\ must be prescribed so that the boundary condition $\rho
(s)\equiv 0$\ on the improper hypersurface of $U_{g}$\ is fulfilled. As a
consequence, in Eq.(\ref{LAST-ONE}) it follows identically that
\begin{equation}
\int_{s_{o}}^{s_{1}}ds\int_{U_{g}}d(g)\frac{\partial \rho (s)}{\partial s}%
\equiv 0,
\end{equation}%
and similarly%
\begin{equation}
\int_{s_{o}}^{s_{1}}ds\int_{U_{g}}d(g)\frac{\partial }{\partial g_{\mu \nu }}%
\left( \rho (s)V_{\mu \nu }(s)\right) \equiv 0,
\end{equation}%
the two equation thus yielding that%
\begin{equation}
Q\left( \psi ,\psi ^{\ast }\right) =\int_{-\infty }^{+\infty
}ds\int_{U_{g}}d(g)\rho (s)\left[ \frac{\partial S^{(q)}(s)}{\partial s}%
+H_{c}^{(q)}\right] .
\end{equation}%
Then, one can show that the Euler-Lagrange equations implied by the quantum
Hamilton variational principle (\ref{HAMILT-VARIATIONAL}), written for the
two real fields $\rho (s)$\ and $S^{(q)}(s)$, namely
\begin{eqnarray}
\frac{\delta Q\left( \psi ,\psi ^{\ast }\right) }{\delta S^{(q)}} &=&0,
\label{EL-1} \\
\frac{\delta Q\left( \psi ,\psi ^{\ast }\right) }{\delta \rho } &=&0,
\label{EL-2}
\end{eqnarray}%
necessarily recover the quantum hydrodynamic equations, i.e., respectively
Eqs.(\ref{cont1}) and (\ref{hj-2}). In fact, first Eq.(\ref{EL-1})
manifestly requires%
\begin{equation}
-\frac{\partial }{\partial s}\rho (s)-\frac{\partial }{\partial g_{\mu \nu }}%
\left( \rho (s)\frac{1}{\alpha L}\frac{\partial S^{(q)}}{\partial g^{\mu \nu
}}\right) =0,
\end{equation}%
which coincides with the quantum continuity equation (\ref{cont1}). Second,
noting that the variation of the Bohm-like potential vanishes identically,
the second Eq.(\ref{EL-2}) recovers exactly Eq.(\ref{hj-2}). Therefore also
in terms of the fluid fields $\left\{ \rho (s),S^{(q)}(s)\right\} $\ the
functional $Q\left( \psi ,\psi ^{\ast }\right) $\ is variational.

Concerning the CQG-quantum Hamilton-Jacobi equation determined here, one
notices that the same equation generalizes the classical GR-Hamilton-Jacobi
equation determined in Ref.\cite{noi5} in the framework of the
manifestly-covariant Hamiltonian theory of GR. As a consequence, Eq.(\ref%
{hj-2}) must imply the validity of corresponding Hamilton equations to be
expressed in terms of the effective quantum Hamiltonian density $H_{c}^{(q)}$%
. Nevertheless, due to the presence of the Bohm-like effective quantum
potential $V_{QM}(g,\widehat{g}(r),r,s),$ the latter now generally must
depend explicitly on the proper time $s$. This contribution is expected to
give rise to proper-time dependent solutions of the non-stationary CQG-wave
equation, while the Bohm-like quantum potential vanishes in the
semiclassical limit prescribed letting $\hslash \rightarrow 0$. This
theoretical feature establishes a logical consistency of the CQG-theory with
the classical Hamilton-Jacobi equation recalled above (see Eq.(\ref{HJ
equation})) and earlier determined in Ref.\cite{noi6}, while at the same
time the connection between CQG-QHE and the corresponding CQG-wave equation
marks a strong analogy of the present quantum theory with the Schr\"{o}%
dinger equation and the generalized Klein-Gordon equation reported in Ref.%
\cite{EPJP-2015} holding for relativistic quantum mechanics. Finally, an
important issue must be mentioned, related to the fluid description
underlying the CQG-wave equation. This concerns a generalization of the
background space-time picture adopted in this work, toward realization of
second-quantization theory, i.e., in which quantum sources are taken into
account. Specifically, this includes quantization and quantum modifications
of the background space-time as well the inclusion of specific possible
quantum particle sources (such as the Hawking radiation, graviton sources,
etc.). The fluid description obtained here can be adopted for investigating
such a route, providing the basis for a statistical description of quantum
gravity theory characterized by second-quantization effects and independence
of geometrical background.

\section{Generalized Heisenberg inequalities for $g-$quantization}

In this section we present proofs of Heisenberg inequalities from first
principles holding for the CQG-wave equation in the framework of $g-$%
quantization. More precisely, the problem is addressed whether, as a
consequence of the strict positivity and smoothness of the quantum PDF $\rho
(s)$ and in analogy with standard quantum mechanics, the quantum state $\psi
(s)$ might/should satisfy suitable Heisenberg inequalities which are related
to the fluctuations (and corresponding standard deviations) of the
Lagrangian variable $g_{\mu \nu }$ and of conjugate quantum canonical
momentum $\pi _{\mu \nu }^{(q)}$, and if the same inequalities might place a
constraint on the proper-time evolution of the quantum state.

For this purpose we determine preliminarily the expectation values and
corresponding fluctuations (i.e. the squared of the standard deviations)
which are associated with the generalized Lagrangian coordinates $g_{\mu \nu
}$. These are respectively prescribed in terms of the expectation values%
\begin{equation}
\widetilde{g}_{\mu \nu }\equiv \left\langle g_{\mu \nu }\right\rangle \equiv
\left\langle \psi |g_{\mu \nu }\psi \right\rangle ,
\label{g-expectation-value}
\end{equation}%
and%
\begin{equation}
\left\langle \left( \Delta g_{\mu \nu }\right) ^{2}\right\rangle \equiv
\left\langle \psi \left\vert \left( g_{\mu \nu }-\widetilde{g}_{\mu \nu
}\right) \left( g_{(\mu )(\nu )}-\widetilde{g}_{(\mu )(\nu )}\right) \right.
\psi \right\rangle .
\end{equation}%
The corresponding weighted configuration-space integrals are then given by
the following expressions:%
\begin{eqnarray}
\widetilde{g}_{\mu \nu } &=&\int_{U_{g}}d(g)\rho g_{\mu \nu },
\label{g-integral-1} \\
\left\langle \left( \Delta g_{\mu \nu }\right) ^{2}\right\rangle
&=&\int_{U_{g}}d(g)\rho \left( g_{\mu \nu }-\widetilde{g}_{\mu \nu }\right)
\left( g_{(\mu )(\nu )}-\widetilde{g}_{(\mu )(\nu )}\right) .
\label{g-integral-2}
\end{eqnarray}%
Similar calculations can be performed for the conjugate quantum momenta,
namely for the expectation values
\begin{equation}
\widetilde{\pi }_{\mu \nu }\equiv \left\langle \pi _{\mu \nu
}^{(q)}\right\rangle \equiv \left\langle \psi |\pi _{\mu \nu }^{(q)}\psi
\right\rangle ,
\end{equation}%
and%
\begin{equation}
\left\langle \left( \Delta \pi _{\mu \nu }^{(q)}\right) ^{2}\right\rangle
\equiv \left\langle \psi \left\vert \left( \pi _{\mu \nu }^{(q)}-\widetilde{%
\pi }_{\mu \nu }\right) \left( \pi _{(\mu )(\nu )}^{\left( q\right) }-%
\widetilde{\pi }_{(\mu )(\nu )}\right) \right. \psi \right\rangle .
\end{equation}%
Straightforward calculations yield in this case:%
\begin{eqnarray}
\widetilde{\pi }_{\mu \nu } &=&\int_{U_{g}}d(g)\psi ^{\ast }\left( -i\hbar
\frac{\partial }{\partial g^{\mu \nu }}\right) \psi  \notag \\
&=&\int_{U_{g}}d(g)\rho \left( \frac{\partial S^{(q)}}{\partial g^{\mu \nu }}%
-\frac{i\hbar }{2}\frac{\partial \ln \rho }{\partial g^{\mu \nu }}\right)
=\int_{U_{g}}d(g)\rho \frac{\partial S^{(q)}}{\partial g^{\mu \nu }},
\label{p-integral-1}
\end{eqnarray}%
and respectively:%
\begin{eqnarray}
\left\langle \left( \Delta \pi _{\mu \nu }^{(q)}\right) ^{2}\right\rangle
&=&\int_{U_{g}}d(g)\psi ^{\ast }\left( -i\hbar \frac{\partial }{\partial
g^{\mu \nu }}-\widetilde{\pi }_{\mu \nu }\right) \left( -i\hbar \frac{%
\partial }{\partial g^{(\mu )(\nu )}}-\widetilde{\pi }_{(\mu )(\nu )}\right)
\psi  \notag \\
&=&\int_{U_{g}}d(g)\rho \left( -i\hbar \frac{\partial \ln \rho }{\partial
g^{\mu \nu }}+\frac{\partial S^{(q)}}{\partial g^{\mu \nu }}-\widetilde{\pi }%
_{\mu \nu }\right) \left( -i\hbar \frac{\partial \ln \rho }{\partial g^{(\mu
)(\nu )}}+\frac{\partial S^{(q)}}{\partial g^{(\mu )(\nu )}}-\widetilde{\pi }%
_{(\mu )(\nu )}\right)  \notag \\
&&+\int_{U_{g}}d(g)\rho \left( -\frac{\hbar ^{2}}{2}\frac{\partial ^{2}\ln
\rho }{\partial g^{(\mu )(\nu )}\partial g^{\mu \nu }}-i\hbar \frac{\partial
^{2}S^{(q)}}{\partial g^{(\mu )(\nu )}\partial g^{\mu \nu }}\right) .
\label{p-integral-2}
\end{eqnarray}%
From the last expression it follows in particular that the fluctuation $%
\left\langle \left( \Delta \pi _{\mu \nu }^{\left( q\right) }\right)
^{2}\right\rangle $ can be represented as%
\begin{equation}
\left\langle \left( \Delta \pi _{\mu \nu }^{\left( q\right) }\right)
^{2}\right\rangle =\left\langle \left( \Delta \pi _{_{\mu \nu }}\right)
^{2}\right\rangle _{1}+\left\langle \left( \Delta \pi _{_{\mu \nu }}\right)
^{2}\right\rangle _{2},
\end{equation}%
with $\left\langle \left( \Delta \pi _{_{\mu \nu }}\right) ^{2}\right\rangle
_{1}$ and $\left\langle \left( \Delta \pi _{_{\mu \nu }}\right)
^{2}\right\rangle _{2}$ denoting respectively the two weighted integrals:%
\begin{eqnarray}
&&\left. \left\langle \left( \Delta \pi _{_{\mu \nu }}\right)
^{2}\right\rangle _{1}=\frac{\hbar ^{2}}{4}\int_{U_{g}}d(g)\rho \frac{%
\partial \ln \rho }{\partial g^{\mu \nu }}\frac{\partial \ln \rho }{\partial
g^{(\mu )(\nu )}},\right. \\
\left\langle \left( \Delta \pi _{_{\mu \nu }}\right) ^{2}\right\rangle _{2}
&=&\int_{U_{g}}d(g)\rho \left( \frac{\partial S^{(q)}}{\partial g^{\mu \nu }}%
-\widetilde{\pi }_{\mu \nu }\right) \left( \frac{\partial S^{(q)}}{\partial
g^{(\mu )(\nu )}}-\widetilde{\pi }_{(\mu )(\nu )}\right) .
\end{eqnarray}

We intend to prove that the following fundamental inequality holds:%
\begin{equation}
\left\langle \left( \Delta g_{_{(\mu )(\nu )}}\right) ^{2}\right\rangle
\left\langle \left( \Delta \pi _{\mu \nu }\right) ^{2}\right\rangle _{1}\geq
\frac{\hbar ^{2}}{4},  \label{HEIS-1}
\end{equation}%
which implies in turn also that
\begin{equation}
\left\langle \left( \Delta g_{_{(\mu )(\nu )}}\right) ^{2}\right\rangle
\left\langle \left( \Delta \pi _{\mu \nu }^{\left( q\right) }\right)
^{2}\right\rangle \geq \frac{\hbar ^{2}}{4}.  \label{HEIS-2}
\end{equation}%
The two inequalities (\ref{HEIS-1}) and (\ref{HEIS-2}) will be referred to
respectively as \emph{first and second Heisenberg inequalities for }$g-$%
\emph{quantization. }The last inequality can also be represented in terms of
the corresponding standard deviations $\sigma _{g_{_{\mu \nu }}}$ and $%
\sigma _{\pi _{\mu \nu }}$, namely%
\begin{equation}
\left\{
\begin{array}{c}
\sigma _{g_{_{\mu \nu }}}\equiv \sqrt{\left\langle \left( \Delta g_{_{(\mu
)(\nu )}}\right) ^{2}\right\rangle } \\
\sigma _{\pi _{\mu \nu }}\equiv \sqrt{\left\langle \left( \Delta \pi _{\mu
\nu }\right) ^{2}\right\rangle }%
\end{array}%
\right. ,  \label{STANDARD DEFIATION p}
\end{equation}%
thus yielding the equivalent Heisenberg inequality for the standard
deviations in the customary formal representation%
\begin{equation}
\sigma _{g_{_{(\mu )(\nu )}}}\sigma _{\pi _{\mu \nu }}\geq \frac{\hbar }{2}.
\label{HEIS-3}
\end{equation}%
The proof of the two inequalities (\ref{HEIS-1}) and (\ref{HEIS-2}) is
analogous to that given in Refs.\cite{Tessarotto2007}. For this purpose one
notices, first, that by construction the conservation of $\rho (s)$ is
warranted by the validity of the quantum continuity equation (\ref{cont1}).
Hence, thanks to the normalization of $\rho (s)$ according to Eq.(\ref%
{normaliz}) which holds for all $s\in I\equiv
\mathbb{R}
$, for arbitrary proper time $s$, integration by parts in the previous
configuration-space integral delivers equivalently also the identity%
\begin{equation}
-\int_{U_{g}}d(g)\rho \left( g_{\mu \nu }-\widetilde{g}_{\mu \nu }\right)
\frac{\partial \ln \rho }{\partial g_{(\mu )(\nu )}}=1.
\end{equation}%
Hence the inequality%
\begin{equation}
\int_{U_{g}}d(g)\rho \sqrt{\left( g_{\mu \nu }-\widetilde{g}_{\mu \nu
}\right) \left( g_{(\mu )(\nu )}-\widetilde{g}_{(\mu )(\nu )}\right) }\sqrt{%
\frac{\partial \ln \rho }{\partial g_{(\mu )(\nu )}}\frac{\partial \ln \rho
}{\partial g_{(\mu )(\nu )}}}\geq 1
\end{equation}%
manifestly must hold identically for arbitrary $s$ too. Next, Schwartz's
inequality delivers%
\begin{equation}
\int_{U_{g}}d(g)\rho \left( g_{\mu \nu }-\widetilde{g}_{\mu \nu }\right)
\left( g_{(\mu )(\nu )}-\widetilde{g}_{(\mu )(\nu )}\right)
\int_{U_{g}}d(g)\rho \frac{\partial \ln \rho }{\partial g_{\mu \nu }}\frac{%
\partial \ln \rho }{\partial g_{_{(\mu )(\nu )}}}\geq 1.
\end{equation}%
Upon multiplying term by term the LHS of the previous equation with respect
to $\frac{\hbar ^{2}}{4}$ an inequality follows which exactly coincides for
arbitrary $s$ with Eq.(\ref{HEIS-1}). Since by construction $\left\langle
\left( \Delta \pi _{_{\mu \nu }}\right) ^{2}\right\rangle _{2}\geq 0$, the
second Heisenberg inequality (\ref{HEIS-2}) necessarily holds too. Such
inequalities hold for arbitrary $s\in I\equiv
\mathbb{R}
.$ Hence, no possible constraint can arise on the dynamical proper-time
evolution wave-function $\psi (s)$ as a consequence of the validity of the
same inequalities.

The physical interpretation, and consequent implications, implied by the
validity of the Heisenberg inequalities (\ref{HEIS-1}) and (\ref{HEIS-2})
(or equivalently (\ref{HEIS-3})) is of crucial importance for the
prescription of quantum measurements in the context of CQG $g-$quantization.
This occurs if, in analogy with the original Heisenberg interpretation (see
Ref.\cite{Tessarotto2007}), the standard deviations $\sigma _{g_{_{\mu \nu
}}}$ and $\sigma _{\pi _{\mu \nu }}$ defined above are interpreted as \emph{%
quantum measurement errors. }Then, the inequality (\ref{HEIS-3}) states the
impossibility of realizing simultaneous quantum measurements for the
mutually-conjugated Lagrangian coordinate and corresponding momentum, so
that both the canonical variables cannot be measured exactly simultaneously,
i.e., with vanishing standard deviations. In other words, the product of the
corresponding fluctuations/standard deviations is necessarily non-zero since
they must satisfy respectively the fundamental inequalities (\ref{HEIS-2})
and (\ref{HEIS-3}).

\section{Conclusions}

In this paper a number of issues have been posed which are intimately
related to quantum gravity and the statistical interpretation of the
corresponding relativistic quantum wave equation. The goal has been to set
them in the framework of the recently developed axiomatic approach to the
quantization of the metric field tensor (referred to here as $g-$%
quantization) achieved in the context of covariant quantum gravity
(CQG-theory). Such a theory relies on the adoption of the principle of
manifest covariance for the formulation of relativistic statistical
mechanics and in particular for the prescription of a quantum field theory
which is appropriate in the case of the gravitational field.

For this purpose, first, we have shown that the CQG-quantum wave equation
laying at the basis of CQG-theory, can be obtained in the framework of a
Hamilton-Jacobi quantization scheme, i.e., based on the construction of the
classical Hamilton-Jacobi equation and the corresponding
manifestly-covariant Hamiltonian structure associated with the Einstein
field equations. This feature is crucial since it permits the identification
of a specific quantum wave equation, i.e., the evolution equation which
advances the quantum state. In the framework of the background space-time
(first-quantization) manifestly-covariant approach adopted here, the latter
has been identified with a single $4-$scalar wave function $\psi $.

The second step has concerned a crucial aspect of CQG-theory and quantum
field theory alike, namely the establishment of a variational formulation
for the related quantum-wave equation.\textbf{\ }In this paper we have shown
that such an approach can be achieved by means of a suitable variational
principle denoted as quantum Hamilton variational principle. Such a
principle\ - as the analog of the classical Hamilton equations associated
with the same Einstein equations - is a synchronous one, i.e., it is
prescribed so that the variation operator leaves invariant the relevant
volume element in the Hamilton variational functional. In addition, by
construction it exhibits also the property of manifest covariance. As such,
it is frame-independent, i.e., it holds for arbitrary coordinate systems.
Finally, the same variational principle is of general validity in the sense
that it is valid for arbitrary choices of the background space-time, namely
arbitrary classical solutions of the Einstein field equations. As a result,
by means of suitable variations of the Hamilton variational functional, the
quantum wave equation as well as the corresponding set of quantum
hydrodynamic equations, obtained after introducing the Madelung
representation for the wave function, are shown to coincide with the
corresponding Euler-Lagrange equations.

Finally, in reference to the problem of quantum measurements for CQG-theory,
the validity of two Heisenberg inequalities, referred to here respectively
as first and second Heisenberg inequalities, has been proved to hold for
arbitrary quantum wave-functions which are solutions of the same CQG-wave
equation. More precisely, in the context of the $g-$quantization considered
here, the same inequalities refer to quantum measurements of the tensor
field $g\equiv \left\{ g_{\mu \nu }\right\} $ and its conjugate quantum
operator $\pi _{\mu \nu }^{(q)}$, the statistical measurement error
estimates being achieved by means of the corresponding standard deviations.

These conclusions are believed to provide theoretical insight in\ the
axiomatic foundations of CQG-theory. It must be stressed that, in the
current formulation, CQG-theory is built upon a first-quantization approach
assuming existence of a generic background metric that characterizes the
geometrical properties of space-time. This is shown to realize a $g-$%
quantization of the canonical state $(g,\pi )$\ which fulfills by
construction the Quantum Unitarity Principle, and consequently the
conservation of quantum probability associated with the wave function $\psi $%
.\ As a consequence, no trans-Planckian effects nor possible information
losses arising at event horizons in black-hole space-times are yet included
either. Nevertheless, as pointed out above, the inclusion of
second-quantization effects, with particular reference to quantum
modifications of the background space-time at Planck length and quantum
particle sources such as Hawking radiation, is in principle possible. In
view of these considerations, the features highlighted here suggest that
CQG-theory may provide fertile theoretical grounds for a variety of
applications in quantum gravity and quantum cosmology to be envisaged in
future works.

\section{Acknowledgments}

Work developed within the research projects of the Czech Science Foundation
GA\v{C}R grant No. 14-07753P (C.C.) and the Albert Einstein Center for
Gravitation and Astrophysics, Czech Science Foundation No. 14-37086G (M.T.).

\end{document}